\documentclass[11pt,twocolumn]{article}
\usepackage[ruled]{algorithm2e}
\usepackage[switch]{lineno}
\usepackage{geometry}
\usepackage{graphicx}  
\usepackage{epstopdf}
\usepackage{microtype}
\usepackage[ngerman,english]{babel}
\usepackage{caption}
\usepackage[T1]{fontenc}
\usepackage[latin1]{inputenc}
\usepackage{sectsty}
\usepackage{color}
\usepackage{relsize}
\usepackage{appendix}
\definecolor{iemcolor}{rgb}{0.16, 0.26, 0.58}

\allsectionsfont{\normalfont\sffamily\bfseries}
\usepackage{amsmath, amssymb, trfsigns,multicol,upgreek,nicefrac,dsfont,bm,cancel,amsbsy}
\usepackage{url}
\usepackage{booktabs}

\usepackage[colorinlistoftodos,prependcaption,textsize=small]{todonotes}
\usepackage{color}
\usepackage{colortbl}
\usepackage{transparent}
\usepackage{hyperref}
\hypersetup{
	colorlinks,
	linkcolor={blue!80!black},
	citecolor={blue!80!black},
	urlcolor={blue!80!black}
}
\addto\extrasenglish{

}

\usepackage{fancyhdr}
\usepackage{lipsum}% just to generate some text
\graphicspath{{./figures/}}

\usepackage{microtype}

\usepackage{caption}
\usepackage{subcaption}

\makeatletter
\def\@maketitle{%
	\newpage
	\null
	\begin{center}%
		\let \footnote \thanks
		{\LARGE \@title \par}%
		\vskip 1.5em%
		{\large
			\lineskip .5em%
			\begin{tabular}[t]{c}%
				\@author
			\end{tabular}\par}%
		\vskip 1em%
		{\large \@date}%
	\end{center}%
	\par
	\vskip 1.5em}
\makeatother

\makeatletter
\def\@maketitle{%
	\newpage
	\null
	\vspace*{-1cm}
	\begin{center}%
		\let \footnote \thanks
		{\LARGE \@title \par}%
		\vskip 1.5em%
		{\large
			\lineskip .5em%
			\begin{tabular}[t]{c}%
				\@author
			\end{tabular}\par}%
		\vskip 1em%
		{\large \@date}%
	\end{center}%
	\par
	\vskip 0.5em}
\makeatother

\title{{\textbf{Perceptual evaluation of listener envelopment \\ using spatial granular synthesis}}}
\author{\normalsize Stefan Riedel and Matthias Frank and Franz Zotter\\ \normalsize Institute of Electronic Music and Acoustics\\ \normalsize University of Music and Performing Arts Graz, Austria}
\date{}

\begin{document}
\selectlanguage{english}
%\linenumbers
% \nolinenumbers
\twocolumn[\begin{@twocolumnfalse}\maketitle\begin{abstract}\noindent% __autocomment__
		Listener envelopment refers to the sensation of being surrounded by sound, either by multiple direct sound events or by a diffuse reverberant sound field. More recently, a specific attribute for the sensation of being covered by sound from elevated directions has been proposed by Sazdov et~al.\ and was termed listener engulfment. This contribution investigates the effect of the temporal and directional density of sound events on listener envelopment and engulfment. A spatial granular synthesis technique is used to precisely control the temporal and directional density of sound events. Experimental results indicate that a directionally uniform distribution of sound events at time intervals $\Delta t < 20$ milliseconds is required to elicit a sensation of diffuse envelopment, whereas longer time intervals lead to localized auditory events. It shows that elevated loudspeaker layers do not increase envelopment, but contribute specifically to listener engulfment. Lowpass-filtered stimuli increase envelopment, but lead to a decreased control over engulfment. The results can be exploited in the technical design and creative application of spatial sound synthesis and reverberation algorithms.
\end{abstract}\vspace*{0.5cm}\end{@twocolumnfalse}]% __autocomment__

\section{Introduction}\label{sec:Intro}

Listener envelopment (LEV) is a perceptual attribute used to characterize the spatial impression of a sound field. It has been investigated by researchers in concert hall acoustics, spatial sound reproduction and electroacoustic music \cite{bradley1995influence, soulodre2003objective, lynch2017perceptual}. According to Berg \cite{berg2009contrasting}, various definitions have been used for envelopment, as the sensation is evoked either by room reverberation or surrounding direct sound events. The unifying factor seems to be the 'sensation of being surrounded by sound' \cite{berg2009contrasting}. This generic definition has been adopted by various researchers in their works \cite{soulodre2003objective, lynch2017perceptual, riedel2022surrounding}. In contrast to apparent source width (ASW), which refers to the horizontal extent of an auditory event, LEV is related to the immersive auditory quality of a scene. It has been shown that LEV strongly correlates with the overall quality of the listening experience \cite{eaton2022subjective}.
\\ Previous work in the field of concert hall acoustics focused on the effect of early/late reverberation and its directional distribution. It suggested that late lateral energy is crucial for listener envelopment \cite{bradley1995objective, bradley1995influence, griesinger1997psychoacoustics}. Studies additionally report correlation with reverberation from front, rear, and overhead directions \cite{morimoto1993new, furuya2001arrival, blauert1986auditory} and the orchestral dynamics \cite{lokki2020perception}. Literature on multichannel sound reproduction studied the required number of loudspeakers and their arrangement to optimally reproduce the spatial impression of a diffuse sound field. It was concluded that as few as four loudspeakers are sufficient in the case of lowpass noise signals or music stimuli, while more loudspeaker directions are required for broadband pink noise signals \cite{hiyama2002minimum, cousins2015subjective, cousins2017effect}.  \\ 
Most of the aforementioned studies varied the number of active loudspeakers and their directional distribution, and the sound stimuli were typically uncorrelated stationary noise signals or reverberated music signals. The required temporal density of sound events to elicit a sensation of diffuse envelopment remains uninvestigated. Literature on the processing lag of the binaural hearing mechanism reports time constants between 50 to 200 milliseconds \cite{blauert1972lag, culling1998measurements, grantham1979detectability}, suggesting that surrounding sound events at significantly shorter time intervals lead to a diffuse and potentially enveloping perception. 

This motivates the following research questions: \\ \textit{What is the required temporal density of surrounding sound events to elicit a sensation of envelopment?} \textit{Is the highest degree of envelopment elicited by a stationary and isotropically diffuse sound field, or by a sound field that exhibits audible spatio-temporal fluctuations / modulations?} These questions are relevant in the technical design and creative application of artificial reverberators \cite{schlecht2015time, alary2021perceptual, hoffbauer2022four, alary2019directional, valimaki2017late}, spatial sound synthesis techniques \cite{lynch2017perceptual, barrett2002spatio, deleflie2009spatial}, and spatial upmixing algorithms \cite{avendano2004frequency, faller2006multiple}. 

Only few experimental studies have been conducted on the spatial impression of 3D versus 2D sound fields, e.g. reporting on perceptual attributes such as  subjective diffuseness \cite{cousins2015subjective, martens2018discrimination}, degree of '3D envelopment' \cite{lee20152d}, or the overall listening experience \cite{eaton2022subjective}. Loudspeaker setups with height layers increased perceived diffuseness over 'ear-height only' arrangements in an experimental study using pink noise signals \cite{cousins2015subjective}. A successive study on perceived diffuseness investigated the effect of the listener's head movements, and showed that 'with-height' reproduction could only enhance auditory diffuseness if listeners were explicitly allowed/asked to tilt their head sideways, effectively moving height loudspeakers into the interaural axis \cite{martens2018discrimination}. To better describe the perceptual effect achieved by height loudspeaker layers, the term listener engulfment (LEG, 'being covered by sound') was proposed by Sazdov et~al.\ \cite{paine2007perceptual}. However, experimental data comparing the perception of envelopment and engulfment is limited \cite{paine2007perceptual, lynch2017perceptual}. 

It seems that further experiments are necessary to clarify the perceptual effects of height layers. Therefore, we additionally address the following research questions: \textit{Do height layers enhance listener envelopment, or rather contribute to a distinct sensation (engulfment)?  How are these attributes affected by the stimulus bandwidth (high-frequency content)?} 

Methodologically, to control the temporal and directional density of sound events in experimental conditions, a spatial granular synthesis technique is employed in this study. Recently, a similar technique used sample-wise assignment of noise signals in order to study perceptual roughness in spatial impulse response rendering and upmixing \cite{meyer2021perceptual}.
Linear time-invariant (LTI), infinite impulse response (IIR) reverberation algorithms such as feedback delay networks (FDN) render an increasingly dense diffuse reverberation tail, which impedes the synthesis of stimuli with well-controlled temporal density of sound events. Nevertheless, recent developments enable directional control \cite{alary2019directional}. LTI finite impulse response (FIR) filters for spatial reverberation \cite{valimaki2017late, valimaki2013perceptual} can be designed to control the spatio-temporal density of a sound field, but the output density will depend on the temporal character of the input signal (transient vs. stationary). An algorithm that directly assigns time-windowed audio signals to specified directions can synthesize both sparse and diffuse sound fields, given a temporally dense or stationary input signal. Therefore, a spatial granular synthesis is the preferred method in this study. 

The first section of the paper describes and evaluates the method of spatial granular synthesis. The second section of the paper describes and discusses listening experiments, which used spatial granular synthesis to investigate the proposed research questions.

\begin{figure}[t]
	\begin{center}
		\includegraphics[width=0.35\textwidth]{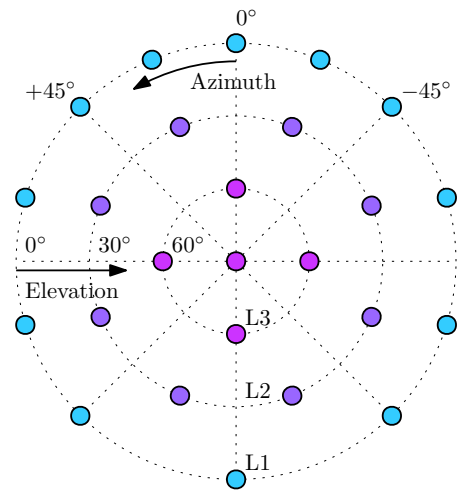}
	\end{center}
	\caption{Schemactic diagram of a hemispherical loudspeaker setup with 25 channels (filled dots). The loudspeakers are grouped into three elevation layers: L1 ($0^{\circ}$ elevation), L2 ($30^{\circ}$ elevation), and L3 ($\geq 60^{\circ}$ elevation).}
	\label{fig:cube_layout}
\end{figure}

\section{Method: Spatial Granular Synthesis}\label{sec:SGS} 
Granular synthesis is rooted in work by Dennis Gabor, who related time-frequency analysis with sonic quanta to human perception of sound \cite{gabor1947acoustical}. Early artistic work with spatialized, layered segments of sound is the piece 'Concret PH' by Iannis Xenakis \cite{xenakis1992formalized}. It was originally presented as an 11-channel tape piece, reproduced via 425 loudspeakers in the Philips Pavilion at Expo 58 \cite{valle2010concrete}. Curtis Roads was the first to implement granular synthesis on digital computer platforms in the 1970s \cite{roads1978automated}, and he explicitly mentions the potential of multichannel granular synthesis in his later works \cite{roads1988introduction, roads2004microsound}. Nuno Fonseca conceptualized and implemented particle systems for audio applications \cite{fonseca2014particle}, where particles can be complete audio files rather than the typically short audio segments used in granular synthesis ($1 \, \mathrm{ms}$ to $200 \, \mathrm{ms}$ \cite{roads2004microsound}). In more recent experimental work, spatialization of audio grains to a frontal array of loudspeakers was used to investigate perceived spatial extent in the horizontal and vertical dimension \cite{weger2016auditory}.  

In spatial granular synthesis, each grain is assigned a position $\bm x = [x,y,z]^\top$ in space, which we might equally express in spherical coordinates $\bm \Omega = (\phi, \theta, r)$ \cite{roads2004microsound, deleflie2009spatial, weger2016auditory, barrett2002spatio}. The possibilities of spatial grain distribution are manifold, e.g.\ we can aim for a directionally uniform distribution around the listener, or restrict the distribution to a specific region in space. 
Several spatial audio techniques can be used to render the grains, e.g. amplitude-panning with loudspeaker arrays \cite{pulkki1997virtual} or virtual sound source positioning with head-related transfer functions (HRTFs) \cite{blauert1997spatial, begault2001direct}. Ambisonics \cite{zotter2019ambisonics} allows encoding of grain objects positioned in a virtual environment, where reproduction takes place via binaural decoding to headphones \cite{zaunschirm2018binaural}, or via decoding to a multichannel loudspeaker system \cite{zotter2012allround}. Lastly, discrete assignment of grains to the nearest available direction in a HRTF database or multichannel loudspeaker arrangement can serve as a baseline method for psychoacoustic experimentation, cf.~Figs.~\ref{fig:cube_layout} and~\ref{fig:sgs_sketch}.

\clearpage

\subsection{Algorithm Definition}\label{subsec:SGS_algo} 

We define a basic spatial granular synthesis algorithm with the following parameters:  
\begin{itemize}
	\setlength\itemsep{0.01em}
	\item $\Delta t$ $\rightarrow$ Time between spatialized grains %(synthesis period)
	\item $L$ $\rightarrow$ Grain length
	\item $w$ $\rightarrow$ Grain window/envelope 
	\item $Q$ $\rightarrow$ Grain seed range in audio buffer
	\item $\bm g$ $\rightarrow$ Weights for spatial rendering
\end{itemize}

\begin{figure*}[hbt]
	\begin{center}
		\includegraphics[width=0.9\textwidth]{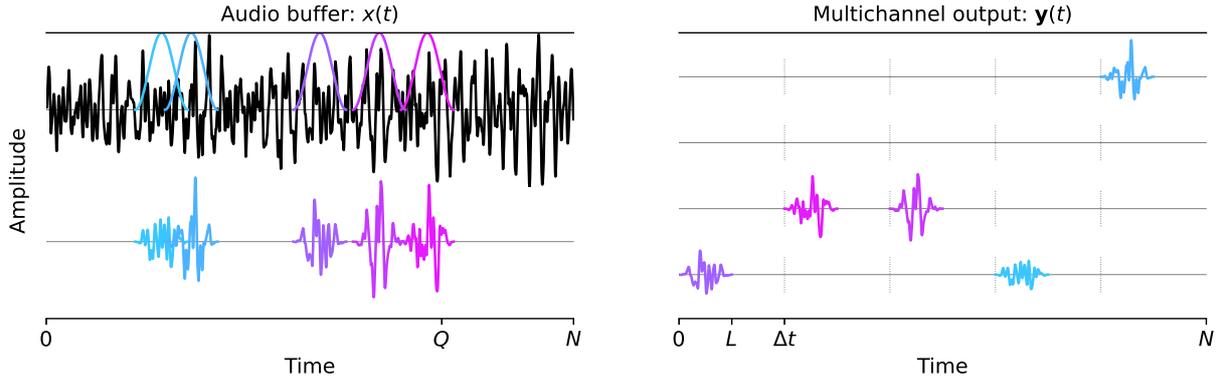}
	\end{center}
	\caption{Spatial granular synthesis extracts grains from an audio buffer and renders them to a multichannel output. The buffer is of length $N$ seconds, and a grain of length $L$ seconds is extracted at time index $q_l$ from the buffer, where $q_l \sim U(0, \, Q )$ in this example. The synthesis period $\Delta t$ is constant at $\Delta t = 2 L$ in this illustration.}
	\label{fig:sgs_sketch}
\end{figure*}

We assume to have access to a signal buffer $x(t)$ of $N > L$ seconds of audio, which can be a recorded sample or a real-time input buffer:
\begin{linenomath*}
	\begin{align}
		x(t) = 
		\begin{cases}
			x(t) \, ,   & \text{for } 0 \leq t \leq N \\
			0, & \text{else} \, .
		\end{cases}
	\end{align}
\end{linenomath*}
The $l$-th grain is extracted at the buffer index $q_l$ and a window function $w(t)$ is applied, to avoid artifacts in the output and shape its timbral properties. For stimulus generation in our experiments we used a Hann window of length $L$ seconds
\begin{linenomath*}
	\begin{align}
		w(t) =
		\begin{cases}
			\mathrm{sin}^2 \left(\frac{t \pi}{L} \right) \, ,   & \text{for } 0 \leq t \leq L \\
			0, & \text{else} \, .
		\end{cases}
	\end{align}
\end{linenomath*}
Spatial rendering of grains to a $J$-channel output $\bm y(t) \in \mathbb{R}^J$ is achieved by multiplication with weights $\bm g_{l}(\bm \Omega_l)  \in \mathbb{R}^J$, which are real-valued in case of discrete assignment, vector-base amplitude panning, and Ambisonics encoding \cite{zotter2012allround}: 
\begin{linenomath*}
	\begin{align}
		\label{eq:sgs_formula}
		\bm y(t) = \frac{1}{\mathcal{G}} \sum_l \bm g_{l} \cdot w(t -  \tau_l) \cdot x(t - \tau_l + q_l) \, ,
	\end{align}
\end{linenomath*}
where the summation considers active grains defined by $0 < (t - \tau_l) < L$ and $\tau_l = l \Delta t$ in case of strictly periodic synthesis. Convolution $(*)$ with filter weights enables direct binaural synthesis using head-related impulse responses $\bm g_l(\bm \Omega_l, \, t) \in \mathbb{R}^2$ ($J=2$):
\begin{linenomath*}
	\begin{align}
		\bm y_\mathrm{LR}(t) = \frac{1}{\mathcal{G}} \sum_l \bm g_{l} * [w(t -  \tau_l) \cdot x(t - \tau_l + q_l)] \, ,
		\label{eq:bin_gran_synth}
	\end{align}
\end{linenomath*}
\begin{linenomath*}
	We ensure constant loudness across varying grain densities by applying a gain factor $1/\mathcal{G}$ that compensates the spatio-temporal grain overlap $\Psi = L / \Delta t$ and the window function:
\end{linenomath*}
\begin{linenomath*}
	\begin{align}
		\mathcal{G} = \sqrt{\frac{L}{\Delta t}} \cdot \sqrt{\frac{1}{L} \int_{0}^{L} w^2(t) dt} \, ,
	\end{align}
\end{linenomath*}

assuming that extracted grains are uncorrelated. This is achieved by modulation or random distribution of the extraction index $q_l \sim U(0, \, Q )$ with $Q \leq  (N - L)$.

%The algorithm can in principle be applied to a real-time monophonic input signal
The algorithm could in principle be applied to any kind of monophonic signal, be it a real-time input signal or an impulse response. A greater amount of variation in the grain extraction index $q_l \sim U(0,Q)$ enhances signal decorrelation in exchange for a longer response ('filter length') and signal displacement. \\ The stimuli in this paper were generated by an (offline) \textit{Python} implementation of the algorithm, which is openly available, see Sec.~\ref{sec:data}. We selected discrete channel-based assignment for the psychoacoustic experiments in this study, to avoid any influence of order-dependent sidelobe levels using Ambisonics or the direction-dependent source widening of VBAP \cite{zotter2012allround,frank2013width}. In case of discrete assignment, the vector $\bm g_l$ has only one non-zero entry to assign a grain to a single target channel, cf.~Fig.~\ref{fig:sgs_sketch}. 

\begin{figure*}[hbt]
	\begin{center}
		\includegraphics[width=1\textwidth]{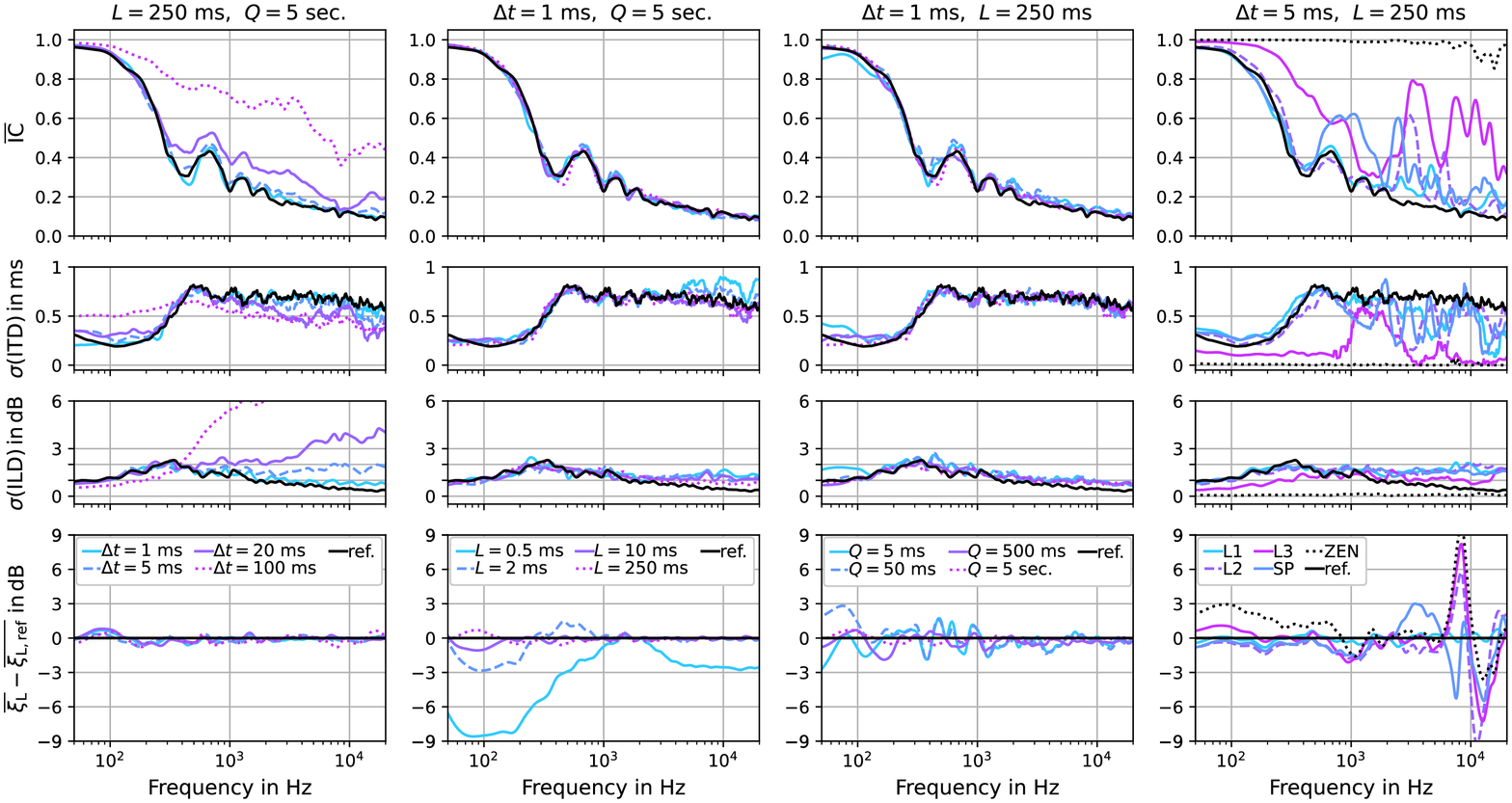}
	\end{center}
	\caption{Auditory cues for spatial granular synthesis versus a 2D diffuse-field reference (black curves). Each column varies a different synthesis parameter (left to right): time between grains $\Delta t$ (column 1), grain lengths $L$ (column 2), seed range $Q$ (column 3), and directional assignment (column 4). Rows show different cues (top to bottom): Mean interaural coherence, standard deviation of ITD, standard deviation of ILD, and monaural difference between mean stimulus spectrum and mean diffuse-field spectrum.  } 
	\label{fig:eval_graphs}
\end{figure*}

\subsection{Relation to FIR Filters}\label{subsec:SGS_relation} 
Formally, the spatial granular synthesis defined in Eq.~\ref{eq:sgs_formula} is equivalent to a \emph{time-variant} FIR system, where the time variation of a filter coefficient $h_l$ follows the grain window as $h_l(t) = h_l \cdot w(t)$. The algorithm reduces to an LTI FIR system, if we replace the time window by unity as $w(t) = 1$, and remove the read offset $q_l$:
\begin{linenomath*}
	\begin{align}
		\bm y(t) &= \frac{1}{\mathcal{G}} \sum_l \bm g_{l} \cdot h_l \cdot x(t - \tau_l) \, ,
	\end{align}
\end{linenomath*}
where the sparse (discrete) convolution considers the non-zero coefficients $h_l$, and the factor $\mathcal{G} = \sqrt{\sum_l h_l^2}$ accounts for the number of coefficients and their gain.

Depending on the temporal character of the input signal, an LTI FIR approach can be used to control the spatio-temporal density of a stimulus by carefully choosing the density of non-zero filter coefficients. However, the LTI FIR system may only reproduce delayed copies of the input signal, which limits the decorrelation in the multichannel output. To increase decorrelation among the output signals, time-variant FIR filter coefficients could be considered, but the implementation as granular synthesis seems most practical. Similarly, generating sparse output signals from a temporally dense input is straightforward with the granular approach using short windows $w(t)$. 
 
\subsection{Instrumental Evaluation of Auditory Cues}\label{subsec:algorithmeval}

 By analyzing auditory cues such as the interaural coherence (IC), interaural time and level differences (ITD and ILD), and monaural spectral cues, we can assess perceptual differences between the synthesized stimuli and a stationary diffuse-field reference \cite{hiyama2002minimum, walther2011assessing}. This allows a more insightful interpretation of the experiment design and results presented in Sec.~\ref{sec:Exp}. Stimulus ear signals $\bm y_\mathrm{LR}(t) = [ x_\mathrm{L}(t), \, x_\mathrm{R}(t)]^\top$ are obtained by convolution of grain objects with (free-field) head-related transfer functions (HRTFs), cf.~Eq.~\ref{eq:bin_gran_synth}. Details on the subsequent computation of the auditory cues can be found in the appendix section. 
 
 A 2D diffuse-field reference is simulated using a circular HRTF set of 360 directions (stationary, uncorrelated pink noise signals at $1^\circ$ azimuth resolution), cf. black curves in Fig.~\ref{fig:eval_graphs}. The simulated stimuli consist of pink noise grains spatialized via convolution with KU100 HRTFs: columns 1 to 3 show uniformly random assignment in the horizontal plane ($1^{\circ}$ resolution), and column 4 shows uniformly random assignment within a direction subset (cf.~Fig.~\ref{fig:cube_layout}).

In Fig.~\ref{fig:eval_graphs} (column 1) the time interval $\Delta t$ between spatialized grains is varied, while the grain length $L$ and seed range $Q$ are constant ($L = 250 \, \mathrm{ms}$, $Q = 5 \, \mathrm{sec.}$). The spatial distribution of grains is uniformly random among 360 circular HRTF directions. At an interval $\Delta t = 100 \, \mathrm{ms}$ (sparse condition), a high IC indicates that pronounced ITDs are extracted by the cross-correlation mechanism, suggesting that the individual sound events are rather localizable. For an interval $\Delta t = 1 \, \mathrm{ms}$, the synthesized sound field resembles a diffuse sound field, as the auditory cues show diffuse-field behaviour (nearly identical to black curves). Low IC corresponds to high fluctuations of the ITDs, such that localization of individual sound events is impeded \cite{faller2004source}, causing a sensation of spaciousness and envelopment. In a diffuse sound field the magnitude of ILD fluctuation is limited to $\sigma(\mathrm{ILD}) \leq 2 \, \mathrm{dB}$, meaning that short-time magnitudes of ILD are slightly above the just-noticeable difference (JND) of ILDs ($\approx 0.5$ to $1 \, \mathrm{dB}$) \cite{hartmann2002interaural}. A reduction of the spatio-temporal density for $\Delta t \geq 5  \, \mathrm{ms}$ causes pronounced ILDs above $1 \, \mathrm{kHz}$.
\\ In Fig.~\ref{fig:eval_graphs} (column 2) the effect of the window length $L$ becomes apparent as a magnitude roll-off around the frequency $f = 1 / L$, which can be seen for short grains of $L \leq 10 \, \mathrm{ms}$. Note that grain lengths $L = 250 \, \mathrm{ms}$ yield an output spectrum that corresponds to the reference condition (spectrum of the input signal).

% Low IC corresponds to high fluctuations of the ITDs, such that localization of individual sound events is impeded \cite{faller2004source}, causing a sensation of spaciousness and envelopment. The magnitude of ILD fluctuation is limited to $\sigma(\mathrm{ILD}) \leq 2 \, \mathrm{dB}$ in a diffuse sound field, meaning that short-time magnitudes of ILD are slightly above the just-noticeable difference (JND) of ILDs ($\approx 0.5$ to $1 \, \mathrm{dB}$) \cite{hartmann2002interaural}. A reduction of the spatio-temporal density for $\Delta t = 20  \, \mathrm{ms}$ causes pronounced ILDs above $1 \, \mathrm{kHz}$, while IC and ITDs remain close to diffuse-field behaviour. This suggests that high-frequency ILDs are robust cues to detect, discriminate and localize individual (short-time) sound events in multi-source conditions  \cite{klockgether2016just, santala2011directional}. 

In column 3 of Fig.~\ref{fig:eval_graphs} the variable parameter is the seed range $Q$ for the selection of grains from the audio buffer. Interestingly, even for values of $Q \ll L$, where phase-correlation between seeded grains tends to increase, the interaural cues still show diffuse-field behaviour. However, spectral peaks and notches become visible in the magnitude spectra for $Q < L$, cf.\ Fig.~\ref{fig:eval_graphs} (column 3), which resemble the comb-filtering behaviour of (correlated) early reflections in rooms. 

%This implies that a real-time spatial diffusion of a monophonic input signal is possible for values $Q \leq 20 \, \mathrm{ms}$, however causing timbral coloration known from e.g. chorus effects or decorrelation filters \cite{kendall1995decorrelation}. To avoid any such coloration effects in the presented listening experiments we chose $Q =  N - L$ and audio buffers of length $N > 5 \, \mathrm{sec.}$ for stimulus generation.

In Fig.~\ref{fig:eval_graphs} (column 4) we evaluate assignment of grains to direction subsets used in the listening experiments. The synthesis parameters are $\Delta t = 5 \, \mathrm{ms}$, $L = 250 \, \mathrm{ms}$,  $Q = 5 \, \mathrm{sec.}$, and directional assignment is uniformly random among the channels of a selected subset: L1, L2, L3, stereophonic (SP, $\pm 45^\circ$ azimuth), or monophonic zenith (ZEN, $90^\circ$ elevation), cf.~Fig.~\ref{fig:cube_layout}.  The horizontal layer L1 yields an IC close to the 2D diffuse-field IC up to around 2 kHz with minor deviations at higher frequencies. The height layer L2 deviates more notably above 2 kHz, and L3 deviates clearly from the diffuse field even below 2 kHz, cf.~Fig.~\ref{fig:eval_graphs} (column 4 top row). The zenith loudspeaker condition (ZEN) is a single-direction stimulus and produces highly correlated ear signals, such that fluctuations in ITD and ILD are absent. Spatialization to height layers yields pronounced spectral features above 6 kHz (pinna cues), cf.~Fig.~\ref{fig:eval_graphs} (column 4 bottom). Most notably, we see a prominent energy peak at $8 \, \mathrm{kHz}$, which is known to be a cue for sound source elevation, cf.\ Blauert's 'directional bands' \cite{rajendran2019spectral, blauert1997spatial}.

\section{Listening Experiments}\label{sec:Exp}

\begin{figure*}[hbt]
	\begin{center}
		\includegraphics[width=\textwidth]{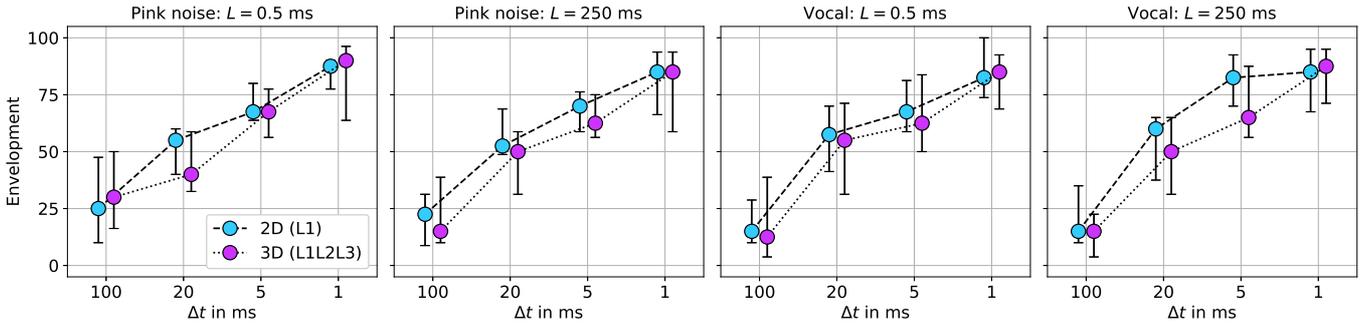}
	\end{center}
	\caption{Median responses and interquartile range ($15$ participants) for trials 1 to 4 of epxeriment I. For 2D conditions grains were randomly assigned to the ear-height loudspeakers (L1), whereas for the 3D conditions grains were assigned randomly among the total set of loudspeakers in the hemisphere (L1L2L3).}
	\label{fig:results_tempdensity}
\end{figure*}

In this section we present two listening experiments. The first experiment investigates the effect of the temporal and directional density of sound events (grains) on envelopment, and the second experiment compares the perception of envelopment and engulfment using various height loudspeaker layers. The participants were given the following definitions of the attributes:
%\begin{samepage}
\begin{itemize}
	\item Envelopment (LEV): being surrounded by sound,
	\item Engulfment (LEG): being covered by sound from above.
\end{itemize}
%\end{samepage}

All participants completed the first experiment on envelopment without knowing that an additional attribute (engulfment) would be defined in the second experiment, in order to avoid bias in their ratings on envelopment regarding 2D vs.\ 3D ('with-height') conditions. Furthermore, none of the presented experiments made use of a reference condition, to avoid any assumptions on what conditions are most enveloping / engulfing.

\subsection{Experiment I: Effect of temporal and directional density on LEV}
\subsubsection*{Setup and Design}
The experiment was conducted at the IEM CUBE, an academic reproduction studio/venue with a reverberation time of $\mathrm{RT_{30}} = 0.5 \, \mathrm{s}$. The hemispherical layout consists of 25 full-range, point-source loudspeakers by \textit{d\&b audiotechnik} and is shown in Fig.~\ref{fig:cube_layout}. The loudspeakers of the setup were individually equalized by 512-taps minimum-phase FIR filters to the mean loudspeaker response in third-octave bands, including frequency-independent gain factors that compensated the level differences as measured from the (double-octave smoothed) frequency responses. 

During the experiment, the listeners were seated centrally. Their head orientation was not constrained, aiming for a natural listening situation as in a concert or installation. \\ The experiment used a multiple stimulus paradigm, however without  a reference, in order to avoid predefining any type of sound field to be most enveloping. Each trial contained 8 conditions of two seconds duration, designed to range from non-enveloping to potentially enveloping scenes. Participants rated the perceived envelopment of the 8 stimuli on a continuous scale from 0 to 100 (0: not at all, 50: moderate, 100: full), presented via a graphical application on a laptop computer. \\
The stimuli were generated by the spatial granular synthesis algorithm described in Sec.~\ref{sec:SGS}, where the algorithm extracts Hann-windowed grains from random positions in the audio input file and assigns them randomly (uniform distribution) to channels of a designated loudspeaker subset, cf.~Fig.~\ref{fig:cube_layout}. The audio buffer was large enough ($N > 5$ sec., $Q> 5$ sec.) to avoid any spectral effects of sampling critically short buffers, cf.~Fig.~\ref{fig:eval_graphs} (column 3 bottom).

\begin{table}[t]
	\tabcolsep12.5pt
	\caption{Bonferroni-Holm corrected $p$-values for three pairwise Wilcoxon signed-rank tests between $\Delta t$ conditions of experiment I. Pink noise (PN) and vocal (VO) trials for each of the spatializations (2D/3D) and grains lengths (0.5 ms and 250 ms) are shown. Bold numbers indicate $p<0.05$. }
	\label{table:pvalues_exp1}
	{%
		\begin{tabular}{@{}lccc@{}}
			$\Delta t $ (ms) & 100 vs. 20  & 20 vs. 5 & 5 vs. 1\\
			\midrule
			2D PN 0.5 ms  &       \textbf{0.005} &     \textbf{0.007} &    \textbf{0.024} \\
			2D PN 250 ms  &       \textbf{0.006} &     0.068 &    \textbf{0.009} \\
			2D VO 0.5 ms  &       \textbf{0.000} &     \textbf{0.024} &    0.051 \\
			2D VO 250 ms  &       \textbf{0.004} &     \textbf{0.003} &    0.959 \\
			3D PN 0.5 ms  &       \textbf{0.013} &     0.055 &    \textbf{0.024} \\
			3D PN 250 ms  &       \textbf{0.008} &     \textbf{0.041} &    \textbf{0.033} \\
			3D VO 0.5 ms &       \textbf{0.016} &     0.060 &    0.060 \\
			3D VO 250 ms &       \textbf{0.000} &     \textbf{0.004} &    \textbf{0.013} \\ 
	\end{tabular}}
\end{table}

\begin{table}[t]
	\tabcolsep12pt
	\caption{Bonferroni-Holm corrected $p$-values for four pairwise Wilcoxon signed-rank tests between 2D and 3D conditions of experiment I. Pink noise (PN) and vocal (VO) trials for each of the grains lengths (0.5 ms and 250 ms) are shown. Bold numbers indicate $p<0.05$.}
	\label{table:pvalues_exp1b}
	{%
		\begin{tabular}{@{}lcccc@{}}
			$\Delta t $  & 100 ms  & 20 ms & 5 ms &  1 ms\\ 
			\midrule
			PN 0.5 ms &  1.000 &  0.428 &  0.464 &  1.000 \\
			PN 250 ms &  1.000 &  0.187 &  1.000 &  1.000 \\
			VO 0.5 ms &  1.000 &  1.000 &  1.000 &  1.000 \\
			VO 250 ms &  \textbf{0.028} &  0.460 &  \textbf{0.021} &  0.875 \\ 
	\end{tabular}}
\end{table}

The trials 1 to 4 used grains of length $L \in \{0.5, 250 \}$ milliseconds extracted from a sound sample of either pink noise (trials 1+2) or a vocal quartet (EBU SQAM Track 48, trials 3+4). Within the trials, $\Delta t$ was varied between $\Delta t \in \{ 100,20,5,1 \}$ milliseconds and the directional assignment was varied between 2D (L1, ear-height loudspeakers) and 3D (L1L2L3, hemisphere). For the impulsive grains ($L = 0.5 \, \mathrm{ms}$) no spatio-temporal overlap occurs, as even for the smallest $\Delta t = 1 \, \mathrm{ms}$ we have $\Psi = L / \Delta t < 1$. \\ A fifth experimental trial was designed to vary only the directional density by restricting grain assignment to one of the following loudspeaker subsets:  stereophonic (SP), quadraphonic (QP), 2D (L1), or 3D (combined layers L1L2L3). The loudspeaker signals of trial 5 were created by $L=250 \, \mathrm{ms}$ Hann-windowed grains assigned randomly within the respective subset every $\Delta t = 1 \, \mathrm{ms}$ ($\Psi = 250$). Due to the high spatio-temporal grain overlap, the loudspeaker signals can be assumed to be approximately stationary noise signals in this trial. As a second independent variable in trial 5, the grains were extracted from either a pink noise or lowpass-filtered pink noise sample ($12^\mathrm{th}$-order Butterworth with a cut-off frequency of 1.8 kHz). \\ Across all trials 1 to 5, the time between grains $\Delta t$ was subject to controlled jitter, limited to 1\% of $\Delta t$, in order to prevent signal periodicity. However, the inherent timbral effects of the window length are likely more relevant, cf.~Fig.~\ref{fig:eval_graphs} (column 2 bottom).

\subsubsection*{Results}
\begin{figure}[t]
	\begin{center}
		\includegraphics[width=0.35\textwidth]{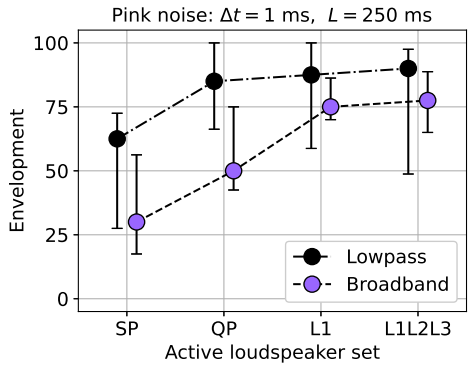}
	\end{center}
	\caption{Median and interquartile range for trial 5 of experiment I. The conditions correspond to stereophonic (SP, $\pm 45^\circ$), quadraphonic (QP, $\pm 45^\circ  \pm 135^\circ$), 2D ear-height layer (L1), and 3D hemisphere (L1L2L3).}
	\label{fig:results_direcdensity}
\end{figure}

Fifteen participants took part in the experiment, either staff or students of the authors' institution. The experimental results of trials 1 to 4 are shown in Fig.~\ref{fig:results_tempdensity}. Per trial, two independent variables were tested, namely the time $\Delta t$ between spatialized grains and the type of spatialization (2D vs.\ 3D). To test the effect of $\Delta t$, we conducted pairwise Wilcoxon signed-rank tests between neighbouring steps of $\Delta t$, and report $p$-values for the 2D and 3D spatialization variants in Tab.~\ref{table:pvalues_exp1}. It turns out that most steps in $\Delta t$ yield statistically significant differences in envelopment. For trial 4 (vocal grains of 250 ms length), the last step from $\Delta t = 5 \, \mathrm{ms}$  to  $\Delta t = 1  \, \mathrm{ms}$ is significant for 3D spatialization, but not for 2D, where ratings reach saturation for $\Delta t = 5 \, \mathrm{ms}$. To test the effect of 2D vs.\ 3D spatialization, pairwise Wilcoxon signed-rank tests were conducted, and $p$-values are reported in Tab.~\ref{table:pvalues_exp1b}. As seen in Tab.~\ref{table:pvalues_exp1b}, the spatialization type does not lead to a significant difference for most conditions. However, in trial 4 at $\Delta t = 5$ ms, where the 2D spatialization apparently reached saturation, the corresponding 3D condition was rated significantly lower ($p = 0.021$).  
%TODO: Show that effect of time interval $\Delta t$ on envelopment is highly significant, and that the effect of 2D/3D is not. We tested 2D against 3D per condition as individual statistical tests, as each spatio-temporal density represents a separate "experiment".
\\The results of trial 5 are shown in Fig.~\ref{fig:results_direcdensity}. Wilcoxon signed-rank tests were conducted within the two signal groups (broadband and 1.8 kHz lowpass pink noise), cf.~Tab.~\ref{table:pvalues_exp1_bw}. A significant difference is found between stereophonic (SP) and quadraphonic (QP) reproduction, for both broadband (unfiltered) and lowpass pink noise ($p=0.003$ and $p=0.007$). Interestingly, there is a significant difference between the quadraphonic (QP) and the L1 condition for broadband pink noise ($p=0.013$), while there is no significant difference between those conditions for the 1.8 kHz lowpass pink noise ($p=0.944$). Between the L1 (2D) and L1L2L3 (3D) conditions, no significant difference can be found, neither for broadband nor for lowpass pink noise signals.

%Table

\begin{table}[t]
	\tabcolsep6.5pt
	\caption{Bonferroni-Holm corrected $p$-values for three pairwise Wilcoxon signed-rank tests between stereophonic (SP), quadraphonic (QP), horizontal layer L1, and hemispherical L1L2L3 conditions of trial 5 of experiment I. Bold numbers indicate $p<0.05$.}
	\label{table:pvalues_exp1_bw}
	{%
		\begin{tabular}{@{}lccc@{}}
			&  SP vs. QP &  QP vs. L1 &  L1 vs. L1L2L3 \\ 
			\midrule
			Lowpass   &             \textbf{0.007} &        0.944 &          0.530 \\
			Broadband &             \textbf{0.003} &        \textbf{0.013}  &          0.752 \\ 
	\end{tabular}}
\end{table}

\subsubsection*{Discussion}
The results in Fig.~\ref{fig:results_tempdensity} indicate that surrounding sound events at an interval of $\Delta t \leq 20 \, \mathrm{ms}$ evoke a moderate to high sensation of envelopment. Even in conditions without any spatio-temporal overlap of sounds ($L=0.5$ ms), the perception becomes diffuse due to the processing lag of the human auditory system \cite{blauert1972lag}. The perceptual integration time $T$ must be greater than 20 ms, as a sensation of envelopment is formed for $\Delta t \leq 20 \, \mathrm{ms}$. On the other hand, an upper bound for the integration time could be estimated as $T < 200 \,\mathrm{ms}$, because envelopment ratings for $\Delta t = 100 \, \mathrm{ms}$ are low as seen in Fig.~\ref{fig:results_tempdensity}. This suggests that individual grains are rather localizable, well resolved auditory events at such a rate of spatialization. It is therefore conclusive to assume a perceptual integration time of $20 \, \mathrm{ms} < T < 200 \, \mathrm{ms}$, which is consistent with literature on the 'binaural sluggishness' of the auditory system \cite{blauert1972lag, grantham1979detectability, culling1998measurements}. The physical overlap $\Psi = L / \Delta t$ does not appear to be a suitable indicator of perception, which is shown by the conditions with grain lengths $L= 0.5 \, \mathrm{ms}$, which all give $\Psi < 1$, but show large perceptual variations from non-enveloping ($\Delta t = 100$ ms) to highly enveloping ($\Delta t = 1$ ms). \\ In contrast, auditory cues such as IC and ILD fluctuations evaluated in Fig.~\ref{fig:eval_graphs} (column 1) plausibly explain the ratings for the different intervals $\Delta t$, given a perceptually meaningful window length is employed ($\mathcal{T} \approx T$). Note that ILD fluctuations are especially pronounced at high frequencies above 2 kHz, cf.~Fig.~\ref{fig:eval_graphs} (column 1), possibly explaining differences between ratings for pink noise and vocal sound stimuli (cf. slopes of the 2D, $L = 250 \, \mathrm{ms}$ stimuli in Fig.~\ref{fig:results_tempdensity}). \\ Interestingly, the effect of 2D vs. 3D spatialization seems to be negligible, with a tendency that grain assignment to the 2D loudspeaker subset is more effective in producing envelopment. For a fixed $\Delta t$, 3D conditions have grains spread around the full hemisphere, leaving the more effective horizontal layer with sparser signals. This could explain the trend towards lower ratings for 3D at $\Delta t = 20 \, \mathrm{ms}$ and $\Delta t = 5 \, \mathrm{ms}$. At an interval of $\Delta t = 1 \, \mathrm{ms}$ the ratings saturate for both 3D and 2D spatialization of grains.

\begin{figure*}[hbt]
	\begin{center}
		\includegraphics[width=\textwidth]{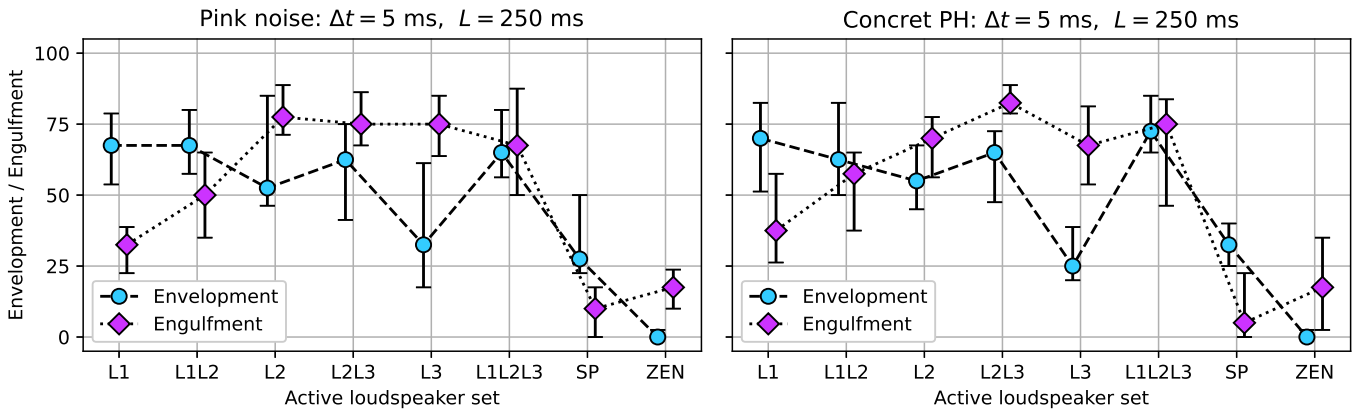}
	\end{center}
	\caption{Median and interquartile range ($15$ participants) for layer-based trials of experiment IIa. Envelopment and engulfment was tested in separate trials, where the independent variable was the loudspeaker subset: L1 (horizontal loudspeakers), L2 (loudspeakers at $30^\circ$ elevation), L3 (loudspeakers at $\geq 60^\circ$ elevation). Hidden anchor conditions were included, namely a stereophonic condition (SP) and a monophonic zenith condition (ZEN, $90^\circ$ elevation).}
	\label{fig:layers_lev_leg}
\end{figure*}

The results of trial 5 agree with previous work on envelopment, which showed that for lowpass noise signals or reverberated music signals, 4 loudspeakers are perceptually close to a 24-loudspeaker (2D) reference  \cite{hiyama2002minimum, riedel2022surrounding}. This is plausible as literature states that discrimination between a directionally sparse loudspeaker setup and a directionally dense reference is more difficult for lowpass noise signals \cite{santala2011directional}. Removing high-frequency signal content prohibits access to certain localization cues, especially high-frequency ILDs (and ITDs), cf. stereophonic reproduction (SP) in Fig~\ref{fig:eval_graphs} (column 4), likely causing the reduced localizability and increased envelopment.  Additionally, the results in Fig.~\ref{fig:results_direcdensity} indicate that 2D spatialization (L1) is able to fully saturate perceived envelopment (even for broadband pink noise).

\subsection{Experiment II: Effect of height loudspeakers and signal bandwidth on LEV vs. LEG}

\begin{figure*}[t]
	\centering
	\begin{subfigure}{0.8\textwidth}
		\centering
		\includegraphics[width=\textwidth]{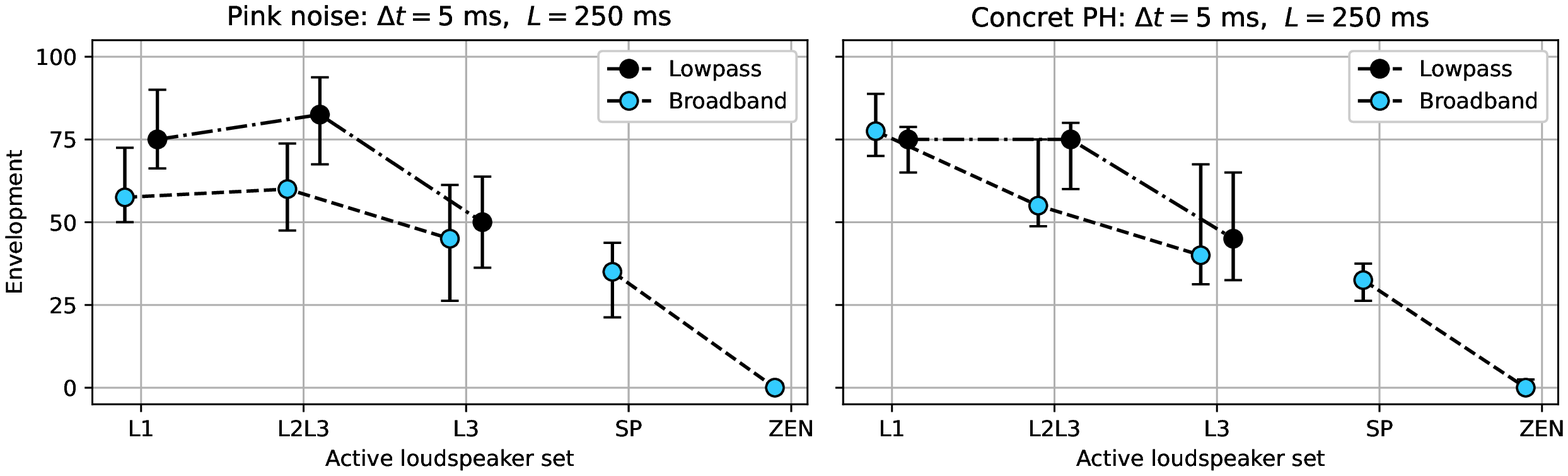}
	\end{subfigure}
	\hfill
	\begin{subfigure}{0.8\textwidth}
		\centering
		\vspace*{-0.0cm}
		\includegraphics[width=\textwidth]{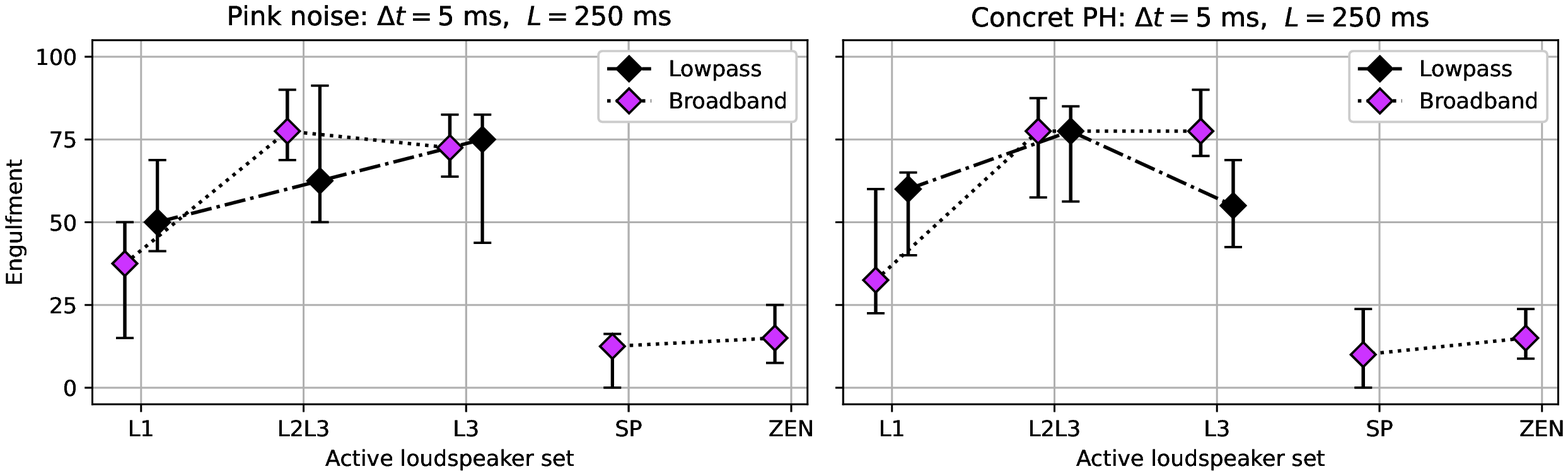}
	\end{subfigure}
	\vspace*{-0.2cm}
	\caption{Median and interquartile range ($15$ participants) for layer- and bandwidth-based trials of experiment IIb. Effect of loudspeaker set on perceived envelopment (top) and engulfment (bottom) for broadband (unfiltered) and lowpass-filtered stimuli (12th-order Butterworth with cut-off at 1.8 kHz). Tested layers are L1 ($0^\circ$ elevation), L2L3 ($\geq 30^\circ$ elevation), and L3 ($\geq 60^\circ$ elevation). Anchor conditions are stereophonic (SP) and a monophonic zenith condition (ZEN).} 
	\label{fig:bandwidths}
\end{figure*}

% format for Heading-B style
\subsubsection*{Setup and Design}
The setup of the second experiment was equivalent to the first experiment, and all participants completed the second experiment after the first experiment. In addition to envelopment, a second attribute called engulfment was introduced to the participants. It is defined as 'the sensation of being covered by sound from above', while the definition of envelopment was repeated as 'the sensation of being surrounded by sound' \cite{lynch2017perceptual}. \\ The experiment was divided into two parts, according to the perceptual attributes envelopment and engulfment. In each part, participants rated only one of the attributes, and the order of the two parts was randomized. Each trial presented 8 stimuli, either varying exclusively the active loudspeaker set (IIa), or varying both the signal bandwidth and the active loudspeaker set (IIb). The monophonic input to the spatial granular synthesis was either pink noise or an excerpt of the composition 'Concret PH' by Iannis Xenakis. For the trials of type IIb, the unfiltered signals were rated against their 1.8 kHz lowpass-filtered versions (12-th order Butterworth). The length of the grains was $L=250 \, \mathrm{ms}$, spatialized at an inverval $\Delta t = 5 \, \mathrm{ms}$, both for the pink noise stimuli and the 'Concret PH' stimuli. This gives a spatio-temporal density of sound events which allows for a moderate to high sensation of envelopment (and supposedly engulfment), cf.\ results of experiment I. The loudspeaker sets defined for the experiment are the $0^\circ$ elevation layer  (L1), the $30^\circ$  elevation layer (L2), and the group of remaining loudspeakers at $60^\circ$ elevation plus the zenith loudspeaker (L3), cf.~Fig.\ref{fig:cube_layout}. A combination of the sets is denoted by concatenation of the abbreviations, e.g.\ L1L2 refers to the combined set of loudspeakers in the L1 and L2 layers. Lastly, two anchor stimuli were provided: a stereophonic condition (SP, $\pm 45^\circ$ azimuth, $0^\circ$ elevation) and a monophonic zenith condition (ZEN, $90^\circ$ elevation).

\begin{table}[t]
	\tabcolsep9.5pt
	\caption{Bonferroni-Holm corrected $p$-values for two pairwise Wilcoxon signed-rank tests between layer conditions of experiment IIa. Bold numbers indicate $p<0.05$.}
	\label{table:pvalues_exp2_levleg_layers}
	{%
		\begin{tabular}{@{}lcccc@{}}
			&\multicolumn{2}{l}{Envelopment}&\multicolumn{2}{l}{Engulfment} \\
			%\midrule
			&   L2L3 &     L3 &   L2L3 &     L3 \\
			\midrule
			L1 (Pink)      &  0.169 &   \textbf{0.021} &  \textbf{0.005 } &  \textbf{0.004} \\
			L1 (Concret PH) &  0.330 &   \textbf{0.005} &  \textbf{0.009}&  \textbf{0.026} \\ 
	\end{tabular}}
\end{table}

\subsubsection*{Results}
Figures~\ref{fig:layers_lev_leg}  and ~\ref{fig:bandwidths} show results of the second listening experiment. Fig.~\ref{fig:layers_lev_leg} (left) shows the effect of the active loudspeaker layer on envelopment and engulfment for the pink noise stimuli. The L1 condition (ear-height surround) obtained high ratings for envelopment, which is consistent with the results obtained in experiment I. When comparing L1 with the L2L3 and L3 conditions, we find significantly lower envelopment ratings for the L3 condition ($p = 0.021$), but not for the L2L3 condition ($p=0.169$), cf. Tab~\ref{table:pvalues_exp2_levleg_layers} (left). The reported $p$-values result from pairwise Wilcoxon signed-rank tests between L1 and the two other conditions. The results for the 'Concret PH' stimuli in Fig.~\ref{fig:layers_lev_leg} (right) show the same behaviour, where the envelopment rating of L1 is high, and ratings are significantly lower for L3 ($p = 0.005$). \\ \newline
Regarding engulfment, the L1 condition was rated low for pink noise stimuli, cf.\ Fig.~\ref{fig:layers_lev_leg} (left), which is expected for a condition composed of horizontal-only, broadband sound. While the rating of the L1L2 condition is higher than L1, engulfment further increased for conditions purely composed of the height layers L2 and L3, e.g.\ the difference between L1 and L3 shows to be highly significant in terms of engulfment ($p=0.004$), cf.\ Tab.~\ref{table:pvalues_exp2_levleg_layers} (right). Notice that the monophonic zenith loudspeaker condition was rated lower than the horizontal L1 condition in terms of engulfment. \\The results for the 'Concret PH' stimuli show the same trends, cf.\ Fig.\ref{fig:layers_lev_leg} (right). The difference in engulfment between L1 and L3 is also significant ($p=0.026$), and the L2L3 condition achieved the highest rating regarding engulfment. \\ Note that the L2L3 and L1L2L3 conditions achieved high ratings for envelopment \textit{and} engulfment, for both pink noise and 'Concret PH' stimuli.

\begin{table}[t]
	\tabcolsep8pt
	\caption{Bonferroni-Holm corrected $p$-values for two pairwise Wilcoxon signed-rank tests between layer conditions of experiment IIb: Broadband (Bb.) and lowpass (Lp.) signals. Bold numbers indicate $p<0.05$.}
	\label{table:pvalues_exp2_levleg_bandwidths}
	{%
		\begin{tabular}{@{}lcccc@{}}
			&\multicolumn{2}{l}{Envelopment}&\multicolumn{2}{l}{Engulfment} \\
			%\midrule
			&   L2L3 &     L3 &   L2L3 &     L3 \\
			\midrule
			L1 (Bb. Pink)      &  0.804 &  0.127 &  \textbf{0.001} &  \textbf{0.002} \\
			L1 (Lp. Pink)      &  0.400 &  \textbf{0.001} &  0.136 &  0.151 \\
			L1 (Bb. Concret PH) &  \textbf{0.026} &  \textbf{0.011} & \textbf{0.009} &  \textbf{0.010} \\
			L1 (Lp. Concret PH) &  0.656 &  \textbf{0.001} &  \textbf{0.007} &  0.679 \\	
	\end{tabular}}
\end{table}

The results in Fig.~\ref{fig:bandwidths} show the effect of 1.8 kHz lowpass-filtered stimuli on envelopment and engulfment. Lowpass-filtered pink noise stimuli yield a relative increase in envelopment over broadband stimuli for the L1 and L2L3 conditions, cf.~Fig.~\ref{fig:bandwidths} (top left). The reduction in envelopment from L1 to L3 is visible for all signal types, and interestingly turns out to be significant for the lowpass-filtered pink noise stimuli ($p=0.001$) and the lowpass-filtered 'Concret PH' stimuli ($p=0.001$), cf.~Tab.~\ref{table:pvalues_exp2_levleg_bandwidths} (left). Regarding engulfment, ratings of the horizontal L1 conditions increase for lowpass-filtered stimuli, cf.~Fig.~\ref{fig:bandwidths} (bottom). Certain height-layer conditions show reduced engulfment ratings for the lowpass-filtered stimuli, such as L2L3 in case of the pink noise stimuli and L3 in case of the 'Concret PH' stimuli. \\ The difference in engulfment is significant between the L1 condition and the height conditions L2L3 ($p=0.001$) and L3  ($p=0.002$) for the broadband pink noise stimuli, while it is not significant for the lowpass-filtered pink noise stimuli, cf.\ Tab.~\ref{table:pvalues_exp2_levleg_bandwidths} (right). Similarly, for the 'Concret PH' stimuli the difference is significant between L1 and L3 for broadband (unfiltered) stimuli, but not so for the lowpass-filtered stimuli.

%and potentially decreased ratings for the L2L3 and L3 conditions. While there is a higly significant difference between L1 and and L2L3 for the broadband stimuli ($p<0.01$, \underline{better give exact p values here in text}), there is no significant difference for the lowpass-filtered stimuli. Between L1 and L3, the difference is highly significant for broadband pink noise, and significant for lowpass-filtered pink noise ($p<0.05$). This indicates that lowpass-filtered stimuli suffered from decreased localizability and cause a more diffuse perception of the soundfield. 

\subsubsection*{Discussion}
The results of experiment II confirm that envelopment and engulfment are distinct perceptual attributes, as initially proposed by Sazdov et al.\ \cite{paine2007perceptual}. They are controllable by varying the active loudspeaker layer, which is especially clear when comparing the ratings for the L1 and the L3 conditions. While the L1 condition was rated as highly enveloping, it obtained low ratings for engulfment. Contrarily, the L3 condition was rated as engulfing but delivered a low sensation of envelopment. This is plausible, as the L3 condition did not supply any direct sound from directions below elevation $\theta = 60^\circ$, causing a lack of surrounding auditory events. However, the L3 conditions provide height localization cues (8 kHz peak) and a sufficiently low interaural coherence, cf.~Fig.~\ref{fig:eval_graphs} (column 4), which seems to be the psychoacoustic foundation for engulfment. The zenith condition (ZEN) obtained low ratings for envelopment and engulfment, which is explained by the high interaural coherence ($\mathrm{IC} \approx 1$), corresponding to an absence of fluctuations in ITDs and ILDs. Engulfment clearly cannot be achieved by a single elevated sound source. 
%A recent study \cite{lynch2017perceptual}, which used identical definitions for envelopment and engulfment, could reveal a significant difference in perceived envelopment varying the active loudspeaker layer, but could not reveal the same for engulfment. We must assume it was due to the particular experiment design, e.g. the loudspeaker setup, room acoustics, or the spatialization techniques used in the respective experiment.  \\ 
\\ The second part of experiment II investigated the effect of signal bandwidth and active loudspeaker set. Regarding engulfment, lowpass-filtered stimuli reduced the difference between the horizontal L1 condition and the height conditions (L2L3 and L3), cf.~Fig.~\ref{fig:bandwidths} (bottom). This could be due to the localization uncertainty (blur) introduced by the 1.8 kHz lowpass-stimuli, which especially affects localization in the median plane, relying on monaural spectral cues above 2 kHz \cite{baumgartner2014modeling}. Although engulfment is controlled more stably with broadband stimuli, our results demonstrate that some lowpass-filtered stimuli were perceived as engulfing, which can be explained by binaural height localization cues available for laterally elevated sounds (lateral vertical planes) \cite{butler1992localization}. Butler and Humanski \cite{butler1992localization} showed that vertical localization with 3.0 kHz lowpass stimuli fails in the median plane, but is functional in a lateral vertical plane. These effects together with dynamic listening cues available through head movements could explain the higher variance in the engulfment ratings for the lowpass-filtered stimuli, compared to the more clear separation of conditions for broadband stimuli, cf.\ Fig.~\ref{fig:bandwidths} (bottom).

\newpage
\section{Conclusion}
We proposed spatial granular synthesis as a method to generate sound fields with variable temporal and directional density. Listening experiments were conducted in a hemispherical loudspeaker array, and results indicate that listener envelopment requires surrounding sound events at intervals $\Delta t < 20$ milliseconds. Reduction of the time interval between sound events showed a monotonic increase of listener envelopment in our experiments, where saturation is reached for perceptually diffuse sound fields.

If multiple surrounding sound events occur within a sufficiently short time frame $T$, they cannot be individually resolved and localized. The auditory event becomes perceptually diffuse and enveloping, even when no simultaneous directional overlap was present. The perceptual integration time was found to be $20 \, \mathrm{ms} < T < 200 \, \mathrm{ms}$, which corresponds to literature on the binaural processing lag of the auditory system \cite{blauert1972lag, culling1998measurements, grantham1979detectability}. A running analysis of interaural coherence could explain the experimental responses regarding temporal density, provided that the temporal analysis window is consistent with the perceptual integration time (e.g. $\mathcal T \approx 100$ ms).
%This was confirmed by a model based on interaural cross-correlation, which shows optimal predictions for integration times of $43  \,  \mathrm{ms}  \leq T_\mathrm{iacc} \leq 85 \, \mathrm{ms}$. 
\\  Additionally, the design of the experiments did not make suggestions whether 3D (with-height) or 2D conditions would deliver a better sensation of envelopment and/or engulfment. This allowed us to show that the ear-height loudspeaker layer contributes most effectively to envelopment, whereas height loudspeaker layers contribute primarily to engulfment. This can be explained by the fact that height layers provide monaural spectral cues for vertical localization, but can lead to an increase in the interaural coherence depending on their elevation level.
 \\ Our results demonstrate a reduced control over engulfment for 1.8 kHz lowpass-filtered stimuli. High-frequency signal content shows to be beneficial to control the sensation of engulfment, as it relies on perceptual cues for height localization, e.g. monaural spectral cues.  In contrast, the sensation of envelopment can be enhanced by lowpass stimuli. The 1.8 kHz lowpass-filtered pink noise increased envelopment especially for directionally sparse conditions (2 or 4 active loudspeaker directions) due to the increased localization blur. \\  It should be noted that 3D (with-height) loudspeaker systems can render sound fields that will be perceived as both enveloping and engulfing (cf. L1L2L3 condition). This might be the cause why envelopment and engulfment were often understood as one sensation of '3D envelopment' or 'subjective auditory diffuseness' \cite{eaton2022subjective, cousins2015subjective}. Our results underline that detailed studies should treat both sensations separately.  %Broadband pink noise signals require a dense directional coverage to elicit a sensation of envelopment. 
 %This underscores the distinct psychoacoustic nature of the two perceptual attributes.

%\vspace{-0.2cm}
\section{Open Data and Software}
%\vspace{-0.2cm}
\label{sec:data}
As the stimulus generation method was so effective and versatile as a tool, it was worth the effort to implement an open-source, Ambisonic granular synthesis VST plug-in available online at \url{https://plugins.iem.at}.
We provide access to experiment data and code \cite{stefan_riedel_2022_7342635} and to binaural auralizations of the experiment stimuli \cite{stefan_riedel_2022_7342614}. 
%A real-time implementation called 'GranularEncoder' is made available at \url{https://plugins.iem.at}, as spatial granular synthesis proved an effective method for spatial sound diffusion.

\section{Acknowledgment}
This work was partly funded by the Austrian Science Fund (FWF): P 35254-N.

% - If you are using bibTex for references, you need these lines: 
\vspace*{-0.5cm}

%\break

%Appendix

\clearpage

\appendix

\section*{Appendix: Computation of Auditory Cues}\label{sec:algorithmeval}

%\section{Evaluation of Auditory Cues}\label{sec:algorithmeval}

%In this section the spatial granular synthesis algorithm is evaluated regarding its ability to render a perceptually diffuse sound field. By analyzing auditory cues such as the interaural coherence (IC), interaural time and level differences (ITD and ILD), and monaural spectral cues, we can assess perceptual differences between synthesized stimuli and a diffuse-field reference \cite{hiyama2002minimum, walther2011assessing}. \\ The required ear signals $\bm y_\mathrm{LR}(t) = [ x_\mathrm{L}(t), \, x_\mathrm{R}(t)]^\top$ are obtained by convolution of grain objects with (free-field) head-related transfer functions (HRTFs), cf.~Eq.~\ref{eq:bin_gran_synth}. 
Our analysis of auditory cues is time- and frequency-dependent as in the auditory system ('running spectral analysis' \cite{blauert1986auditory}). The temporal window of the binaural hearing mechanism \cite{culling1998measurements} is modeled by analyzing sequential signal blocks of $\mathcal{T}$ milliseconds. The frequency resolution of the auditory system is modeled by a bank of (zero-phase) gammatone frequency windows $w_b(\omega)$  \cite{hohmann2002frequency}, where $b$ is the frequency band index and $\omega$ denotes the radial frequency. The short-time Fourier transforms of the ear signals are denoted as $X_\mathrm{L}(\omega, \, t) = \mathcal{F}\{ x_\mathrm{L}(t) \}_\mathcal{T}$ and $X_\mathrm{R}(\omega, \, t) = \mathcal{F}\{ x_\mathrm{R}(t) \}_\mathcal{T}$, where $\mathcal{T}$ denotes the analysis block length. The following equations represent computations per signal block, and therefore omit the notation of time dependence. \\ Interaural coherence (IC) is computed as the normalized, maximum absolute value of the interaural cross-correlation function $R_\mathrm{LR}[b, \tau]$ \cite{blauert1986auditory}:
\begin{linenomath*}
	\begin{align}
		\mathrm{IC}[b] &= \frac{\underset{\tau}{\mathrm{max}} \left|  R_\mathrm{LR}[b, \tau] \, \right|}{\sqrt{P_\mathrm{L}[b] \cdot P_\mathrm{R}[b]}} \, ,
		\label{eq:IC_model}
	\end{align}
\end{linenomath*}
\begin{linenomath*}
	\begin{align}
		R_\mathrm{LR}[b, \tau] &= \int_{- \infty}^{\infty} w^2_b(\omega) X^*_\mathrm{L}(\omega) X_\mathrm{R}(\omega) \, e^{j \omega \tau}\, d \omega \, , \\ 
		P_\mathrm{L}[b] &= \int_{- \infty}^{\infty} w^2_b(\omega) |X_\mathrm{L}(\omega)|^2 \, d \omega \, , \\ 
		P_\mathrm{R}[b] &= \int_{- \infty}^{\infty} w^2_b(\omega) |X_\mathrm{R}(\omega)|^2 \, d \omega \, ,
		\label{eq:R}
	\end{align}
\end{linenomath*}

where the search range for the lag $\tau$ is typically limited to $-1 \, \mathrm{ms} \leq \tau \leq 1 \,  \mathrm{ms}$, and $(^*)$ denotes complex conjugation.
The interaural time and level differences (ITD and ILD) are computed as \cite{katz2014comparative}:
\begin{linenomath*}
	\begin{align}
		\label{eq:ITD_ILD_model}  
		\mathrm{ITD}[b] &=  \underset{\tau}{\mathrm{argmax}} \, R_\mathrm{LR}[b, \tau] \, , \\
		\mathrm{ILD}[b] &=  10 \cdot \mathrm{log}_{10} \left( \frac{P_\mathrm{L}[b]}{ P_\mathrm{R}[b]} \right)\, \mathrm{dB}  \, .  
	\end{align}
\end{linenomath*}
To investigate monaural ear signal spectra we compute:
\begin{linenomath*}
	\begin{align}
		\xi_\mathrm{L}[b] &=  10 \cdot \mathrm{log}_{10} (P_\mathrm{L}[b])\, \mathrm{dB}, \\
		\xi_\mathrm{R}[b] &=  10 \cdot \mathrm{log}_{10} (P_\mathrm{R}[b])  \, \mathrm{dB} .
		\label{eq:spectrum_model}  
	\end{align}
\end{linenomath*}

%\begin{figure*}[hbt]
%	\begin{center}
%		\includegraphics[width=1\textwidth]{GranularEvaluationPlot.pdf}
%	\end{center}
%	\caption{Evaluation of auditory cues (defined in Sec.~\ref{sec:algorithmeval}) for spatial granular synthesis stimuli versus a 2D diffuse-field reference (black curves). Each column varies a different synthesis parameter (left to right): time between grains $\Delta t$ (column 1), grain lengths $L$ (column 2), seed range $Q$ (column 3), and directional assignment (column 4). Pink noise grains are spatialized via convolution with KU100 HRTFs: columns 1 to 3 show uniformly random assignment in the horizontal plane ($1^{\circ}$ resolution), and column 4 shows uniformly random assignment within a direction subset (cf.~Fig.~\ref{fig:cube_layout}). } %For $L \ll 10$ ms, the effect of the window length becomes visible in the ear signal spectra (right). }
%	\label{fig:eval_graphs}
%\end{figure*}

%Spectra are normalized at 2 kHz before subtraction of the reference.
A pink noise buffer of $N = 10$ seconds duration is the input to the spatial granular synthesis, and the output are binaural signals of five seconds duration, simulated using HRTFs of the Neumann KU100 dummy head \cite{bernschutz2013spherical}. The binaural signals are split into successive blocks of $\mathcal{T}=85 \, \mathrm{ms}$, a perceptually motivated window length \cite{culling1998measurements, riedel2022temporal}, and for each block the equations~\ref{eq:IC_model} to~\ref{eq:spectrum_model} are evaluated for 320 gammatone magnitude windows\footnote{The bandpass windows are spaced on an equivalent rectangular bandwidth (ERB) frequency scale ($1/8$ ERB spacing), and each window covers one ERB \cite{glasberg1990derivation}.} $w_b(\omega)$   \cite{hohmann2002frequency}. 

 Per simulated sound field stimulus, we compute the (temporal) mean $\overline{\mathrm{IC}}$ of the frequency-dependent interaural coherence. Since the interaural time and level differences are zero-mean for a balanced, left-right symmetric distribution of sound events, the standard deviations $\sigma(\mathrm{ITD})$ and $\sigma(\mathrm{ILD})$  are computed to measure the amount of temporal fluctuation \cite{blauert1986auditory}. Lastly, we assess the difference between the mean left-ear spectrum of a stimulus $\overline{\xi_\mathrm{L}}$ and the mean left-ear spectrum of a 2D diffuse-field reference $\overline{\xi_\mathrm{L, ref}}$.

\clearpage

\bibliography{bib}
\bibliographystyle{IEEEtran}

\end{document}